\documentclass[times,twocolumn]{aastex61}

\usepackage{graphicx}
\usepackage{xcolor}
\usepackage{amssymb}
\usepackage{amsmath}
\usepackage{gensymb}
\usepackage{units}
\usepackage{tensor}
\usepackage{enumitem}
\usepackage{savesym}  
\savesymbol{tablenum}
\usepackage{siunitx}
\restoresymbol{SIX}{tablenum}
\usepackage{csvsimple}
\usepackage{booktabs}

\newcommand{\decl}{\delta_{\mathrm{ecl}}}
\newcommand{\dvar}{\delta_{\mathrm{var}}}
\newcommand{\dtr}{\delta_{\mathrm{tr}}}
\newcommand{\dk}{\delta_{k}}
\newcommand{\sdk}{\sigma_{\delta}}
\newcommand{\pok}{\phi_{k}}
\newcommand{\spok}{\sigma_{\phi}}
\newcommand{\ddn}{\delta_{\mathrm{dn}}}
\newcommand{\sdn}{\sigma_{\mathrm{dn}}}
\newcommand{\dew}{\delta_{\mathrm{ew}}}
\newcommand{\sew}{\sigma_{\mathrm{ew}}}

\shorttitle{Phase Offsets and Energy Budgets}
\shortauthors{J. C. Schwartz et al.}

\accepted{in ApJ, 2017 October 20}

\begin{document}
	
\title{Phase Offsets and the Energy Budgets of Hot Jupiters}

\correspondingauthor{Joel C. Schwartz}
\email{joel.schwartz@mail.mcgill.ca}

\author{Joel C. Schwartz}
\altaffiliation{McGill Space Institute; Institute for Research on Exoplanets}
\affil{Department of Earth \& Planetary Sciences, McGill University, 3450 rue University, Montreal, QC, H3A 0E8, CAN}
\affil{Department of Physics, McGill University, 3600 rue University, Montreal, QC, H3A 2T8, CAN}
\affil{Department of Physics \& Astronomy, Northwestern University, 2145 Sheridan Road, Evanston, IL, 60208, USA}

\author{Zane Kashner}
\affil{Department of Mathematics, Stanford University, 450 Serra Mall, Stanford, CA, 94305, USA}

\author{Diana Jovmir}
\altaffiliation{McGill Space Institute; Institute for Research on Exoplanets}
\affil{Department of Physics, McGill University, 3600 rue University, Montreal, QC, H3A 2T8, CAN}
\affil{Department of Mathematics \& Statistics, Universit{\'e} de Montr{\'e}al, 2920 chemin de la Tour, Montreal, QC, H3T 1J4, CAN}

\author{Nicolas B. Cowan}
\altaffiliation{McGill Space Institute; Institute for Research on Exoplanets}
\affil{Department of Earth \& Planetary Sciences, McGill University, 3450 rue University, Montreal, QC, H3A 0E8, CAN}
\affil{Department of Physics, McGill University, 3600 rue University, Montreal, QC, H3A 2T8, CAN}

\begin{abstract}
	Thermal phase curves of short-period planets on circular orbits provide joint constraints on the fraction of incoming energy that is reflected (Bond albedo) and the fraction of absorbed energy radiated by the night hemisphere (heat recirculation efficiency). Many empirical studies of hot Jupiters have implicitly assumed that the dayside is the hottest hemisphere and the nightside is the coldest hemisphere. For a given eclipse depth and phase amplitude, an orbital lag between a planet's peak brightness and its eclipse---a phase offset---implies that planet's nightside emits greater flux. To quantify how phase offsets impact the energy budgets of short-period planets, we compile all infrared observations of the nine planets with multi-band eclipse depths and phase curves. Accounting for phase offsets shifts planets to lower Bond albedo and greater day--night heat transport, usually by $\lesssim 1\sigma$. For WASP-12b, the published phase variations have been analyzed in two different ways, and the inferred energy budget depends sensitively on which analysis one adopts.\ Our fiducial scenario supports a Bond albedo of $0.27^{+0.12}_{-0.13}$, significantly higher than the published optical geometric albedo, and a recirculation efficiency of $0.03^{+0.07}_{-0.02}$, following the trend of larger day--night temperature contrast with greater stellar irradiation. If instead we adopt the alternative analysis, then WASP-12b has a Bond albedo consistent with zero and a much higher recirculation efficiency. To definitively determine the energy budget of WASP-12b, new observational analyses will be necessary.
\end{abstract}


\nocite{cowan2011model} 

\section{Introduction}
\label{sec:intro}
Short-period planets on circular orbits are expected to have permanent day and night hemispheres. Only the dayside absorbs stellar radiation, but if the planet has an atmosphere, then some energy can be moved to the nightside. The process can be described by Bond albedo, $A_{B}\in[0,1]$, the fraction of incident flux the planet reflects, and heat recirculation efficiency, $\varepsilon\in[0,1]$, the fraction of absorbed energy transported from day to night. One can constrain both parameters using the planet's day and night effective temperatures, $T_{d}$ and $T_{n}$.

\begin{figure*}
	\centering
	\includegraphics[width=0.9\linewidth]{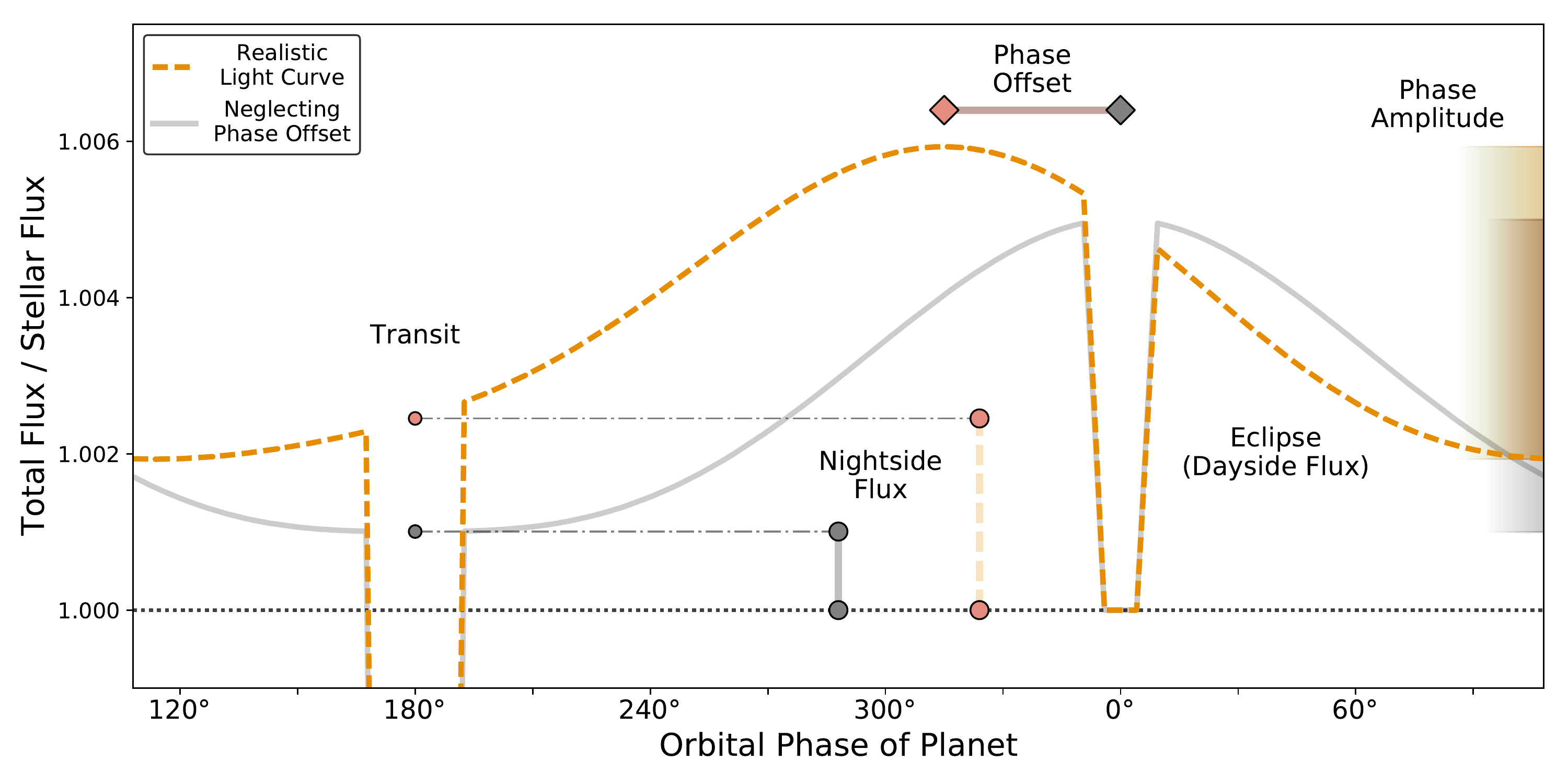}
	\caption{Light curves of a transiting planet with a given eclipse depth and phase amplitude. The horizontal dotted black line denotes the unobscured stellar flux. The dashed orange curve has a non-zero phase offset. If one neglects this offset then one is instead adopting the gray curve, as done in previous energy budget studies \citep[e.g.][]{schwartz2015balancing}. To lowest order, the planet's nightside flux is the eclipse depth minus the peak-to-trough phase amplitude, but this is only exact if the planet exhibits no phase offset. For a fixed eclipse depth and phase amplitude, accounting for a non-zero phase offset (dashed orange curve) will lead one to infer greater nightside flux.}
	\label{fig:Phase_Offset_Diagram}
\end{figure*}

A notional thermal phase curve for a planet is shown with the orange line in Figure~\ref{fig:Phase_Offset_Diagram}. The flux varies because one sees different planetary phases over time, from the nightside at transit to the dayside at eclipse. By combining the eclipse depth, phase variations, transit depth, and stellar spectrum, one can infer the planet's day and night brightness temperatures. By combining brightness temperatures at many wavelengths, one can estimate effective temperatures of a planet's day and night hemispheres.

Many previous studies of hot Jupiters have neglected phase offsets, instead assuming that the dayside is the hottest hemisphere and the nightside is the coolest \citep{cowan2011statistics,perez2013atmospheric,schwartz2015balancing,komacek2017atmospheric}. This is denoted by the gray line in Figure~\ref{fig:Phase_Offset_Diagram}. Since those authors used the actual eclipse depths, the dayside estimates were accurate; but, by adopting the published phase amplitudes and assuming that the nightside was the coolest hemisphere, they underestimated the nightside brightness and hence temperature.

In Section~\ref{sec:data}, we review the compiled data for our energy budget model, including new observations at thermal wavelengths. In Section~\ref{sec:method}, we describe how we use phase offsets in this model, then fit the Bond albedo and recirculation efficiency of nine short-period giant planets. We discuss our results and conclude in Section~\ref{sec:discuss}.

\section{Energy Balance}
\label{sec:EBs}
For transiting planets, one infers Bond albedo and heat recirculation efficiency from infrared observations by estimating effective temperatures for the planet's day and nightsides. This is described by Equations 4--6 of \citet{cowan2011statistics}, which were used and expanded on by \citet{schwartz2015balancing}.

\subsection{Data}
\label{sec:data}
Data for these studies were collected by reviewing published papers and searching both exoplanet.eu \citep{schneider2011defining} and exoplanets.org \citep{han2014exoplanet}. We start with the six planets in Table 2 of \citet{schwartz2015balancing} that have thermal eclipses and phase amplitudes (at wavelengths longward of $0.8~\mu$m): HD~149026b, HD~189733b, HD~209458b, WASP-12b, WASP-18b, and WASP-43b. Then we add WASP-14b \citep{wong2015W14b}, HAT-P-7b, and WASP-19b \citep{wong20163} to our sample. We also incorporate new data from \citet{zhou2015secondary}, \citet{evans2015uniform}, and \citet{line2016no}.
 
We collect first-order phase offsets (Sections~\ref{sec:method} and \ref{sec:2nd_order}) from \citet{knutson20098,knutson2009multiwavelength} and \citet{wong2015W14b,wong20163}. \citet{knutson20098} concluded the offset they found was not statistically significant, so we use their largest uncertainty ($72\degree\pm61\degree$). Phase offsets through \emph{second} order come from \citet{cowan2012thermalW}, \citet{knutson20123}, \citet{maxted2013spitzer}, \citet{zellem20144}, and \citet{stevenson2014thermal,stevenson2017spitzer}. However, the second-order components in \citet{zellem20144} were found to be unnecessary and those at $4.5~\mu$m in \citet{cowan2012thermalW} are disputed, so we use neither in our fits. Also, \citet{cowan2012thermalW} reported two sets of fit parameters for WASP-12b based on different models for detector systematics\footnote{\citet{stevenson2014deciphering} fit eclipse depths for WASP-12b but not phase parameters, so we do not use their values in our analysis.}---we use parameters from their preferred polynomial model but test the other scenario in Section~\ref{sec:W12b}.\footnote{WASP-12 has binary companion stars that affect photometry of the system \citep{bechter2014wasp}. We use dilution factors from \citet{stevenson2014transmission} to correct observations of WASP-12b when appropriate.} All of the nine planets in our study have non-zero offsets in at least one waveband.

\subsection{Model}
\label{sec:method}
We take an energy balance approach to interpreting thermal phase variations \citep[e.g.][]{cowan2011statistics}: we compare the radiation going into and coming out from a planet to infer bulk energetics of that planet's atmosphere.

For our study we use the energy balance model described in Section~3.1 of \citet{schwartz2015balancing}, which accounts for uncertainties in system parameters, as well as reflected light contamination, meridional heat transport, and other sources of uncertainty. Those authors treated stars as blackbodies; we estimate better stellar brightness temperatures for each observation by using BT-NextGen spectra \citep{allard2012models}. Then we use our compiled data to calculate the relative intensity of planets and their host stars at each observed wavelength. In \citet{cowan2011statistics} and \citet{schwartz2015balancing}, this nightside intensity ratio, $\psi_{n}(\lambda)$, is defined as:
\begin{equation}
\label{eq:orig_psin}
\psi_{n}(\lambda) = \frac{\decl - \dvar}{\dtr},
\end{equation}
where $\decl$ is the eclipse depth, $\dvar$ is the peak-to-trough phase amplitude of the full light curve, and $\dtr$ is the transit depth. This is exact only when an observation has no phase offset (e.g. gray curve in Figure~\ref{fig:Phase_Offset_Diagram}).

More generally, one can model the flux $F_{p}$ from a planet as a Fourier series:
\begin{equation}
\label{eq:general_F}
F_{p}(\phi) \approx F_{0} + \sum_{k=1}^{k_{\rm max}} \frac{\dk}{2}\cos\left[k\left(\phi - \pok\right)\right],
\end{equation}
where $\phi$ is the planet's orbital phase ($0\degree$ at eclipse), and $\dk$ and $\pok$ are the phase amplitude and offset of order $k$. Six of the ten published papers with phase offsets use phase curves like this to model their data; we convert parameters from the other studies into the form of Equation~\ref{eq:general_F}. We then modify Equation~\ref{eq:orig_psin} to:
\begin{equation}
\label{eq:new_psin}
\psi_{n}^{\prime}(\lambda) = \frac{\decl - \left[F_{p}(0\degree)-F_{p}(180\degree)\right]}{\dtr}.
\end{equation}
For a given eclipse depth and phase amplitude, this increases the brightness of a planet's nightside when there is an offset and reduces to Equation~\ref{eq:orig_psin} otherwise (cf.\ dashed orange and solid gray curves in Figure~\ref{fig:Phase_Offset_Diagram}). 

If zonal heat advection is the dominant process governing thermal phase curves, then there should be a one-to-one correspondence between the amplitudes and phase offsets of bolometric phase curves. This was first noted by \citet{crossfield2015observations} and is discussed in Appendix~\ref{sec:T_contrast}. In practice, though, we do not yet have bolometric phase curves for any exoplanets: $<50\%$ of the dayside flux and much less of the nightside flux has been captured for most hot Jupiters \citep[Section 2.3 of][]{schwartz2015balancing}. Moreover, we suspect that clouds and magnetic fields might influence hot Jupiter phase curves \citep{Parmentier2016,Rogers2017}. We therefore take the published phase amplitudes and offsets at face value and do not worry about whether they are consistent with the zonal advection hypothesis. However, we do establish that the published \emph{uncertainties} on phase amplitudes and offsets are self-consistent (Appendix~\ref{sec:curve_prec}).

\begin{figure*}
	\centering
	\includegraphics[width=0.9\linewidth]{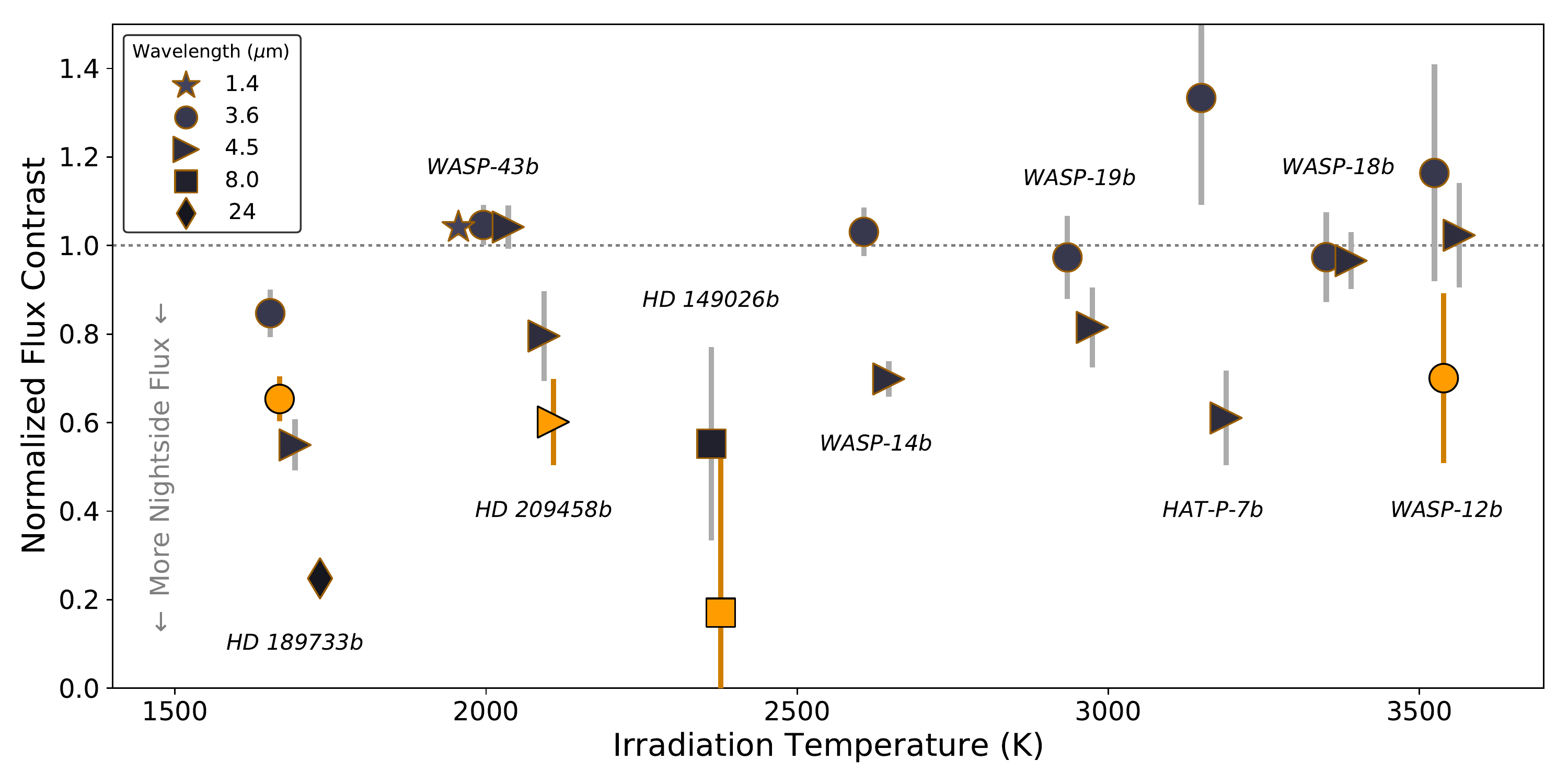}
	\caption{Following \citet{perez2013atmospheric}, we plot normalized day--night flux contrast, $(F_{\textrm{day}}-F_{\textrm{night}})/F_{\textrm{day}}$, versus irradiation temperature, $T_{0} \equiv T_{*}\sqrt{R_{*}/a}$, where $T_{*}$ is the stellar effective temperature, $R_{*}$ is the stellar radius, and $a$ is the planet's semi-major axis. Each marker represents a published observation with a phase offset, and the vertical lines show $1\sigma$ uncertainties (if larger than the marker). The horizontal dotted line shows where the nightside emits no flux. We label the nine planets in our sample and shift markers horizontally for clarity. Dark markers with gray lines show the flux contrasts when neglecting phase offsets \citep{perez2013atmospheric}, while orange symbols show the more accurate contrasts accounting for offsets---we plot the corrected flux contrast if it changes by $\geq0.05$. For most observations, inferred nightside flux only changes a little after including phase offsets, but in a few cases it increases more significantly.}
	\label{fig:17_Flux_DayNight_Diffs}
\end{figure*}

\subsubsection{Second-Order Phase Variations}
\label{sec:2nd_order}
The light curves in Figure~\ref{fig:Phase_Offset_Diagram} are composed of multiple modes, i.e.\ several $\delta_{k}$ are non-zero in Equation~\ref{eq:general_F}. When we use the semi-analytic energy balance model of \citet{cowan2011model} to calculate a planet's flux (Appendices~\ref{sec:T_contrast} and \ref{sec:appar_effec}), the resulting light curves always have non-zero harmonics. We fit these synthetic light curves up to fourth order and find that $\delta_{2}/\delta_{1} \approx $ 0.1--0.2, $\delta_{3}\approx0$ \citep[as expected for edge-on orbits;][]{cowan2013light}, and $\delta_{4}/\delta_{1}\lesssim0.01$. We repeat these fits on light curves of HD~189733b, HD~209458b, and WASP-43b from a global circulation model \citep{zhang2017constraining} and get very similar results. 

Published phase curves have $\delta_{2}/\delta_{1}\lesssim0.15$ \citep[e.g.][]{maxted2013spitzer,stevenson2017spitzer}. Only for the 4.5 $\mu$m phase curve of WASP-12b is the second-order amplitude greater \citep[\mbox{$\sim0.6$;}][]{cowan2012thermalW}, but the authors note it could not be due to planetary temperature patterns so we neglect this second-order component. In any case, we use all published phase amplitude and offset data for Equation~\ref{eq:new_psin} so our results are as robust as possible. First-order phase curves should be accurate to $\sim$15\%, while those reported to second order should be good to $\sim1\%$.

\subsection{The Effect of Phase Offsets}
\label{sec:effect_off}
We reproduce two figures that have previously been used to explore trends in the energy balance of short-period planets, but we now account for phase offsets.

\begin{figure*}
	\centering
	\includegraphics[width=0.85\linewidth]{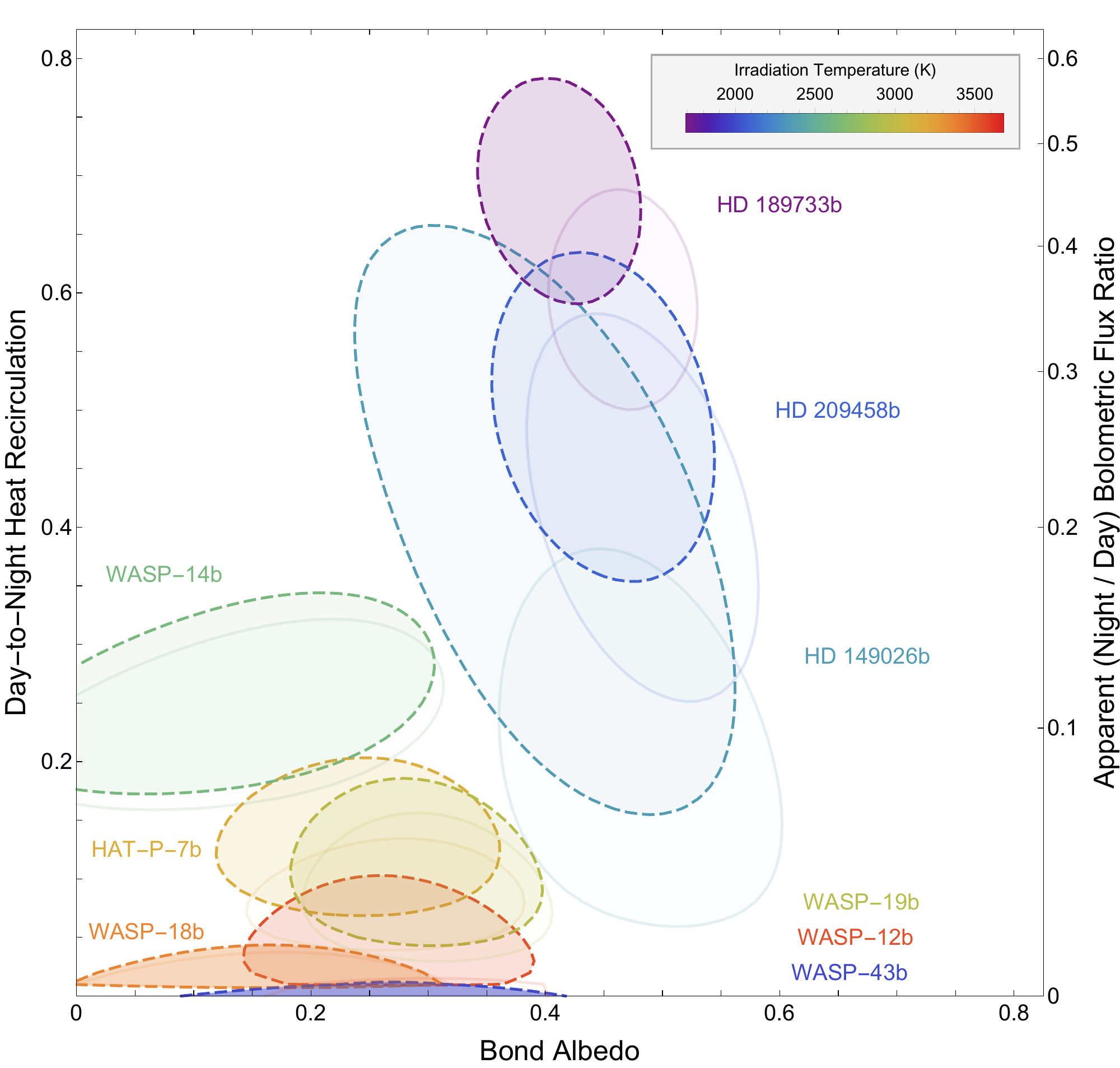}
	\caption{The fitted Bond albedo ($A_{B}$) and day--night heat recirculation efficiency ($\varepsilon$) for nine short-period planets with infrared eclipse and phase data, improved from \citet{schwartz2015balancing}. We also show the apparent bolometric flux ratio for a planet's night and dayside on the right axis (Appendix~\ref{sec:appar_effec}). Light solid curves show the $1\sigma$ regions when neglecting phase offsets, while dashed curves show the more accurate energy budgets accounting for offsets. The color scale shows irradiation temperature; shading is inversely related to the area of the $1\sigma$ region such that tighter fits are darker. The solid curve for WASP-43b is on the bottom axis (its nightside has zero flux if phase offsets are neglected). Every inferred Bond albedo and recirculation efficiency changes by $\lesssim 1\sigma$ when accounting for phase offsets (but see Section~\ref{sec:W12b}).}
	\label{fig:17Spring_Therm_MixedOff_Tight}
\end{figure*}

Figure~\ref{fig:17_Flux_DayNight_Diffs} is similar to Figure~1 from \citet{perez2013atmospheric}. For all infrared observations with a phase offset, we plot the flux contrast when neglecting those offsets as dark markers with $1\sigma$ uncertainties. If accounting for the offset decreases the contrast by at least 0.05, we also show an orange marker. Most nightside fluxes are only modestly affected by phase offsets. But, the $3.6~\mu$m contrasts for HD~189733b and WASP-12b change significantly when phase offsets are accounted for. Since \citet{perez2013atmospheric} ignored phase offsets (their data looks like the gray symbols in our plot), they over-estimated the day--night temperature contrast for those planets.

Figure~\ref{fig:17Spring_Therm_MixedOff_Tight} is similar to Figure~5 of \citet{schwartz2015balancing}. To determine the Bond albedo and recirculation efficiency of each planet, we use either Equation~\ref{eq:orig_psin} or \ref{eq:new_psin} to calculate dayside and nightside brightness temperatures of planets. Next we estimate a planet's day and nightside effective temperatures as the weighted average of its brightness temperatures. We then calculate $\chi^{2}$ on a grid of $A_{B}$ and $\varepsilon$. The $1\sigma$ regions are shown in Figure~\ref{fig:17Spring_Therm_MixedOff_Tight} and colored by irradiation temperature. Following Figure~\ref{fig:Phase_Offset_Diagram}, light solid curves do not account for phase offsets while dashed curves do. We list our fit parameters for the dashed regions in Table~\ref{tab:17_Fitted_PlusOff_Format} and the corresponding changes from the light solid regions in Table~\ref{tab:17_NightAbEp_Diffs}. The dayside and nightside temperatures we report are all \emph{apparent} effective temperatures: the true effective temperatures of the day and night hemispheres are likely lower and higher, respectively, as discussed in Appendix~\ref{sec:appar_effec}.

\section{Discussion and Conclusions}
\label{sec:discuss}
In Figure~\ref{fig:17Spring_Therm_MixedOff_Tight}, nightside temperatures increase toward the upper left. As expected, our fits accounting for phase offsets move to the upper left, with lower Bond albedo and higher day--night heat transport (updated parameters listed in Table~\ref{tab:17_Fitted_PlusOff_Format}). In most cases, the more accurate energy budget constraints agree at the $1\sigma$ level with previous estimates that neglected phase offsets (Table~\ref{tab:17_NightAbEp_Diffs}).\footnote{Although we used second-order phase curves where available, we obtain similar results using only first-order phase curves.}

The exception is the inferred nightside temperature for WASP-12b, which Table~\ref{tab:17_NightAbEp_Diffs} shows is significantly hotter when including phase offsets. Our fitted Bond albedo for WASP-12b is also significantly higher than the planet's optical geometric albedo reported by \citet{bell2017very}.\footnote{\citet{bell2017very} cited the dayside temperature and Bond albedo for WASP-12b from an earlier version of this manuscript.} This is the same tension \citet{schwartz2015balancing} found when analyzing infrared and optical measurements of HD~189733b and HD~209458b.

For WASP-43b, we find that the upper limit on its nightside temperature increases by about an order of magnitude, up from $T_{n}<39$~K at $1\sigma$ when neglecting phase offsets.\footnote{Since the submission of this manuscript, \citet{keating2017revisiting} suggested that the nightside temperature of WASP-43b is in fact in line with HD~209458b.} Besides WASP-12b and WASP-43b, HD~189733b has the most significant changes to its energy budget constraints in Table~\ref{tab:17_NightAbEp_Diffs}.

\subsection{Energy Budget of WASP-12b}
\label{sec:W12b}
The dashed region for WASP-12b in Figure~\ref{fig:17Spring_Therm_MixedOff_Tight} agrees well with Figure~10 from \citet{cowan2012thermalW}. This alone is interesting: we use more eclipse measurements of the planet, correct for binary companion stars diluting those measurements \citep{stevenson2014transmission}, take higher-order phase components into account, and use phase offsets.

As stated in Section~\ref{sec:data}, \citet{cowan2012thermalW} fit their light curves of WASP-12b with two models for detector systematics, polynomial and Gaussian decorrelation. In particular, the polynomial model gives a significantly larger phase amplitude at $4.5~\mu$m, plus a shallower eclipse depth and much larger phase offset ($-53\degree\pm7\degree$ vs. $0\degree\pm29\degree$) at $3.6~\mu$m. We use the polynomial values because the authors argue that their Gaussian decorrelation method removes a lot of the planet's phase signal. But it is hard to decide which systematics model works better using just goodness-of-fit. We therefore fit an alternate energy budget for WASP-12b.

\begin{table}
	\centering
	\caption{Fitted energy budgets accounting for phase offsets.}
	\label{tab:17_Fitted_PlusOff_Format}
	\footnotesize{
		\begin{tabular}{c c c c c}
			\toprule
			Planet & $T_{d} $ (K) & $T_{n}$ (K) & $A_{B}$ & $\varepsilon$\\
			\midrule
			\begin{filecontents*}{t1.csv}
Planet,Td,Tn,AB,eps
HAT-P-7b,$2612\pm93$,$1236\pm178$,$0.25^{+0.11}_{-0.13}$,$0.12^{+0.08}_{-0.05}$
HD 149026b,$1737\pm75$,$1127\pm251$,$0.43^{+0.13}_{-0.19}$,$0.37^{+0.29}_{-0.22}$
HD 189733b,$1163\pm37$,$953\pm44$,$0.41^{+0.07}_{-0.07}$,$0.69^{+0.09}_{-0.1}$
HD 209458b,$1483\pm51$,$1058\pm92$,$0.46^{+0.08}_{-0.11}$,$0.49^{+0.15}_{-0.14}$
WASP-12b,$2939\pm94$,$962\pm354$,$0.27^{+0.12}_{-0.13}$,$0.03^{+0.07}_{-0.02}$
WASP-14b,$2193\pm116$,$1262\pm95$,$0.14^{+0.16}_{-0.14}$,$0.25^{+0.09}_{-0.08}$
WASP-18b,$2905\pm111$,$662\pm378$,$0.16^{+0.15}_{-0.16}$,$0.01^{+0.03}$
WASP-19b,$2407\pm82$,$1069\pm200$,$0.3^{+0.1}_{-0.12}$,$0.1^{+0.09}_{-0.06}$
WASP-43b,$1667\pm56$,$< 430 \ (1\sigma)$,$0.27^{+0.15}_{-0.18}$,$0^{+0.01}$
\end{filecontents*}
\csvreader[head to column names, late after line=\\]{t1.csv}{}{\Planet & \Td & \Tn & \AB & \eps}
			\bottomrule
		\end{tabular}
	}
\end{table}

\begin{table}
	\centering
	\caption{Change in parameters after using phase offsets.}
	\label{tab:17_NightAbEp_Diffs}
	\begin{tabular}{c c c c}
		\toprule
		Planet & $\Delta T_{n}$ (K) & $\Delta A_{B}$ & $\Delta\varepsilon$\\
		\midrule
		\begin{filecontents*}{t2.csv}
Planet,DTn,DAB,Deps
HAT-P-7b,$146\pm246$,$-0.02\pm0.17$,$0.04\pm0.08$
HD 149026b,$214\pm333$,$-0.06\pm0.2$,$0.19\pm0.3$
HD 189733b,$56\pm62$,$-0.06\pm0.09$,$0.1\pm0.13$
HD 209458b,$64\pm157$,$-0.03\pm0.14$,$0.09\pm0.22$
WASP-12b,$646\pm487$,$-0.02\pm0.17$,$0.03\pm0.05$
WASP-14b,$27\pm134$,$0\pm0.22$,$0.02\pm0.12$
WASP-18b,$7\pm516$,$-0.01\pm0.22$,$0\pm0.02$
WASP-19b,$72\pm290$,$-0.01\pm0.15$,$0.02\pm0.1$
WASP-43b,$< 432 \ (1\sigma)$,$0\pm0.23$,$0\pm0.01$
\end{filecontents*}
\csvreader[head to column names,late after line=\\]{t2.csv}{}{\Planet & \DTn & \DAB & \Deps}
		\bottomrule
	\end{tabular}
\end{table}

We repeat our analysis on WASP-12b using parameters from the Gaussian decorrelation model in \citet{cowan2012thermalW} and infer $T_{n}=2190\pm172$~K, a much hotter nightside than in Table~\ref{tab:17_Fitted_PlusOff_Format}. This puts the planet above WASP-14b in Figure~\ref{fig:17Spring_Therm_MixedOff_Tight}, with $A_{B}=0^{+0.1}$ and $\varepsilon=0.52^{+0.13}_{-0.12}$ at $1\sigma$. In this case, WASP-12b does not follow the expectation that hotter planets have much shorter radiative times and so larger day--night contrasts \citep{cowan2011statistics,cowan2012thermalW}, nor theoretical predictions \citep{perez2013atmospheric,komacek2016atmospheric,komacek2017atmospheric}. Instead the Gaussian decorrelation parameters suggest the planet has a moderate recirculation efficiency despite its high irradiation temperature.

If we remain agnostic about which analysis of the extant data is correct, then WASP-12b's nightside temperature is around $1400$--$1800$~K with a Bond albedo of $0.06$--$0.22$ and a heat recirculation efficiency of $0.21$--$0.34$. New phase curves of the WASP-12 system are needed to determine which of the cases above is most accurate.

\section*{Acknowledgments}
The authors thank the anonymous referee for their thorough comments, Kevin B. Stevenson (U. Chicago) for sharing preliminary Spitzer Space Telescope data of \mbox{WASP-43b}, and Emily Rauscher (U. Michigan) for sharing GCM light curves of three planets from an in-prep manuscript. The authors also thank Michael Zhang and Heather Knutson (Caltech) for helpful discussions about and improvements to the energy balance model. JCS acknowledges funding as a Graduate Research Trainee at McGill University. This research has made use of the Exoplanet Orbit Database and the Exoplanet Data Explorer at exoplanets.org.

\bibliographystyle{aasjournal}
\bibliography{JCS_Refs_Thesis_17}

\appendix
\section{A One-to-One Relation Between Phase Offsets and Amplitudes}
\label{sec:T_contrast}
\begin{figure}
	\centering
	\includegraphics[width=0.625\linewidth]{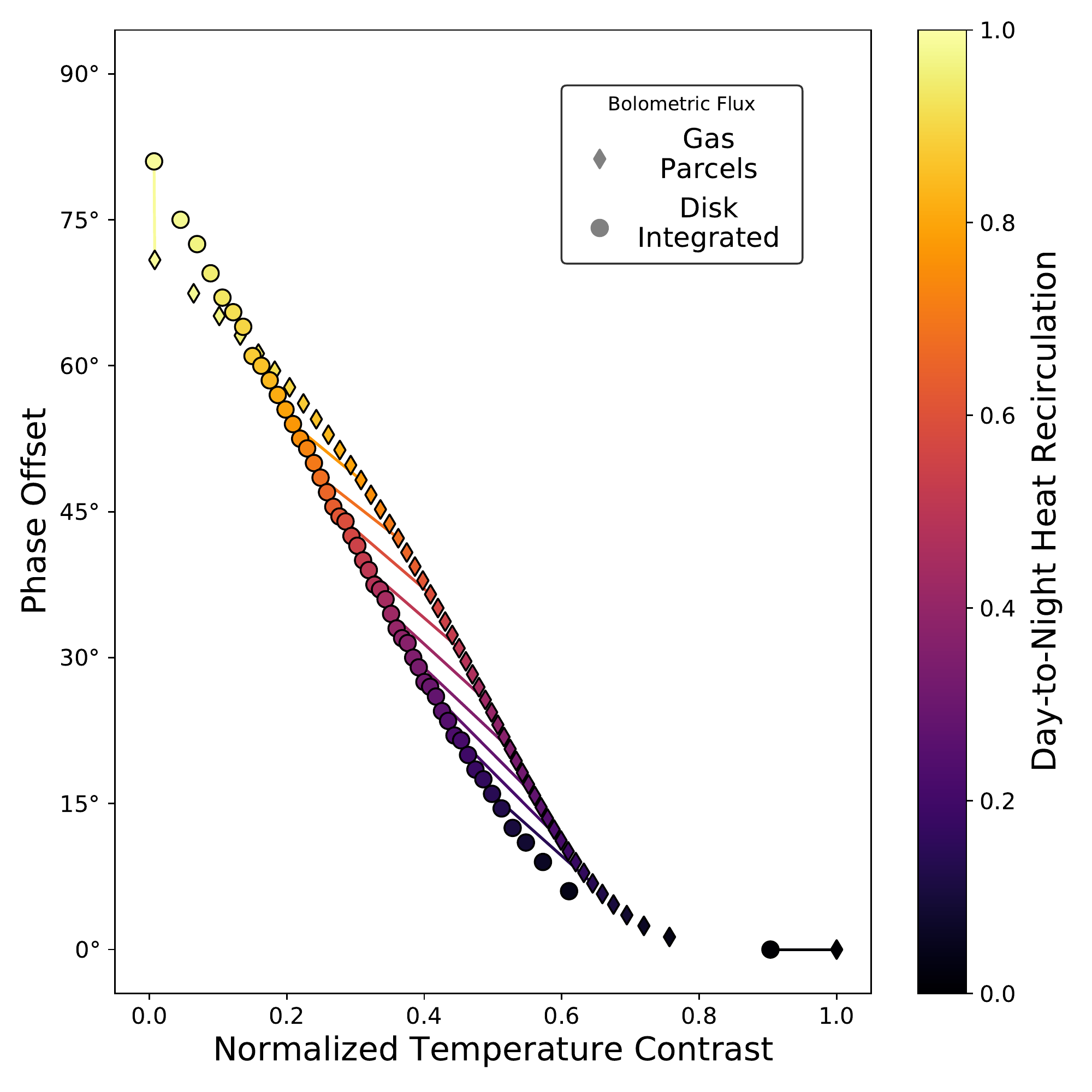}
	\caption{The phase offset versus apparent temperature contrast, \mbox{$(T_{\mathrm{max}} - T_{\mathrm{min}})/T_{0}$,} for recirculation efficiencies in $\sim$0.02 increments (color scale) according to the energy balance model of \citet{cowan2011model}. This is an upgraded version of the left panel of Figure~9 from \citet{crossfield2015observations}, which presented an approximate curve for gas parcels interpolated from only a few cases. Since the one-to-one relation we predict is technically only valid for \emph{bolometric} phase curves, we omit current observational constraints. Both marker types represent bolometric flux: diamonds show values for individual gas parcels, while circles are for disk-integrated flux. Markers are paired by color; we join select pairs with solid lines. In the radiative equilibrium limit (darkest circle, $\varepsilon=0$), the disk-integrated temperature contrast is \mbox{$T_{d}=T_{0}(2/3)^{1/4}$}, as expected. In the limit of efficient zonal heat transport (lightest diamond, $\varepsilon\rightarrow1$), the phase offset of the hot spot approaches \mbox{$\cos^{-1}(1/\pi)\approx71.4\degree$}. Disk integration decreases the temperature contrast, but \emph{increases} the phase offset. This energy balance model predicts a one-to-one correspondence between bolometric phase amplitudes and offsets, which will be tested with upcoming observations that capture a large fraction of the flux from short-period exoplanets.}
	\label{fig:JCS_OffsetvContrast_189_Tight}
\end{figure}

For a planet on a circular orbit with zonal advection of heat, there is a one-to-one relation between bolometric phase offset and bolometric flux contrast because both depend on a single parameter, recirculation efficiency. To demonstrate this, we use the semi-analytic energy balance model of \citet{cowan2011model} and test cases $\sim$0.02 apart in $\varepsilon$ assuming no reflected light (see Appendix~\ref{sec:appar_effec}). For each case, we calculate the phase offset and apparent temperature contrast of the disk-integrated flux, as well as for individual gas parcels, as shown in Figure~\ref{fig:JCS_OffsetvContrast_189_Tight}. Our results agree qualitatively with 3D general circulation models of super-Earths and mini-Neptunes tested by \citet{zhang2017effects}, and they indeed should be universal if heat is zonally advected in the absence of clouds. 

A planet's temperature contrast is strongly affected at very low recirculation (dark markers in Figure~\ref{fig:JCS_OffsetvContrast_189_Tight}) because it only takes a little heat to raise the temperature of a cold nightside. Although the behavior of individual gas parcels (diamonds) and the disk-integrated light curves (circles) are qualitatively similar, there are two quantitative differences. Firstly, disk integration reduces temperature contrast, as expected given the low-pass nature of the convolution \citep{cowan2008inverting,cowan2013light}. But disk integration also \emph{increases} the phase offset: the hottest disk-integrated region of the planet is East of the hottest gas parcel because parcels heat faster than they cool \citep[Figure 1 of][]{cowan2011model}. This means the hot spot is almost never at the center of the planet's brightest hemisphere---one cannot use hot spot offsets and phase offsets interchangeably.

Although the one-to-one correspondence between phase amplitudes and phase offsets was not seen in photometric phase curves \citep{crossfield2015observations}, future missions that measure a greater fraction of the thermal emission from short-period planets should allow us to test this prediction directly (e.g.\ James Webb Space Telescope, FINESSE). As noted by \citet{crossfield2015observations}, deviations from this one-to-one relation would suggest that additional physics---clouds, magnetic fields, etc.---are shaping hot Jupiter phase curves \citep{agundez2012impact,perez2013atmospheric, rauscher2013three,showman2013atmospheric}.

\section{Phase Curve Precision}
\label{sec:curve_prec}
Energy budget estimates like those in Table~\ref{tab:17_Fitted_PlusOff_Format} are only as accurate as the phase amplitudes and offsets that go in them. At a particular order (recall Equation~\ref{eq:general_F}), one can express the phase curve component $F_{k}$ as the sum of a cosine and sine:
\begin{equation}
\label{eq:F_duo}
F_{k} \propto \ddn\cos k\phi + \dew\sin k\phi,
\end{equation}
where $\ddn$ is the day-to-night phase amplitude and $\dew$ is the East-to-West amplitude (we drop the $k$ subscripts for clarity). The two amplitudes are independent variables by Fourier analysis, and their measured uncertainties, $\sdn$ and $\sew$, should be similar for full-orbit observations of phase curves. We test this by fitting toy models of phase curves with Equation~\ref{eq:F_duo} and find that our amplitude uncertainties are generally within a factor of 2.5. For the published phase curves that used the parametrization above, $\sdn$ and $\sew$ differ by $16$--$63\%$ in \citet{knutson20123}, $<4\%$ in \citet{maxted2013spitzer}, and $29\%$ in \citet{zellem20144}, all of which are reasonable.

Alternatively, the flux component of a planetary system can be parametrized by a cosine with a phase offset:
\begin{equation}
\label{eq:F_uni}
F_{k} \propto \dk \cos\left[k\left(\phi - \pok\right)\right].
\end{equation}
This parametrization has the unfortunate property that the uncertainty on $\pok$ diverges when $\dk$ is small. This makes it non-trivial to determine whether a given phase amplitude and offset are appropriately precise.\footnote{This is the same reason it is better to fit for $e\cos\omega$ and $e\sin\omega$ instead of $e$ and $\omega$ with radial velocity data.} The majority of published studies parameterize planetary flux this way.

\begin{figure}
	\centering
	\includegraphics[width=0.6\linewidth]{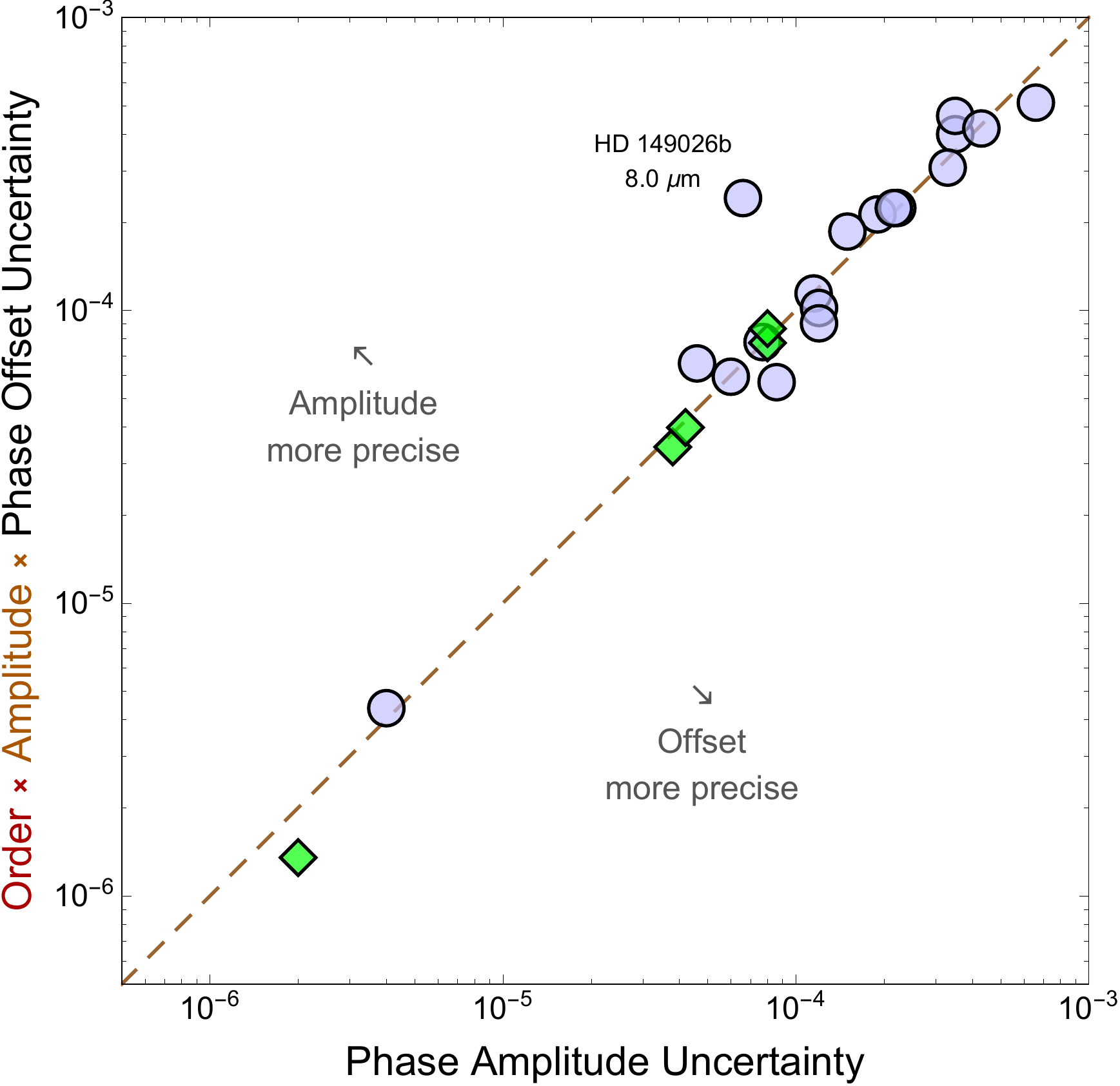}
	\caption{The scaled phase offset uncertainty ($k\dk\spok$) versus the phase amplitude uncertainty ($\sdk$), motivated by Equation~\ref{eq:sigs_relate}. Each marker is a measurement of one planet at one infrared wavelength, where blue circles and green diamonds are first- and second-order data, respectively. Both axes have units of flux and the dashed line shows where these quantities are equal, as expected for sensible $\sdk$ and $\spok$. Only HD~149026b at $8.0~\mu$m has values differing by more than a factor of 2.5, due to a partial phase curve that did not capture the peak flux. Nearly all published phase curves have reasonable phase amplitude and phase offset uncertainties.}
	\label{fig:17Spring_Amp_Off_Unc}
\end{figure}

In order to evaluate the uncertainties $\sdn$ and $\sew$, we differentiate the identities $\ddn = \dk\cos k\pok$ and $\dew = \dk\sin k\pok$, and rearrange them to obtain:
\begin{equation}
\label{eq:sigs_relate}
\sdn^{2} - \sew^{2} = \left[\sdk^{2} - \left(k\dk\spok\right)^{2}\right]\cos 2k\pok,
\end{equation}
where $\sdk$ and $\spok$ are the uncertainties on $\dk$ and $\pok$, respectively.

Because we expect its left-hand side to be close to zero, Equation~\ref{eq:sigs_relate} suggests that we compare $\sdk$ to the product $k\dk\spok$ for our compiled data, shown in Figure~\ref{fig:17Spring_Amp_Off_Unc}. These quantities are reasonable for almost every published observation (i.e.\ the markers are close to the dashed line). The only outlier is HD~149026b at $8.0~\mu$m \citep{knutson20098}, 
but this is unsurprising since their half-orbit phase curve did not capture the phase curve's peak, so the reported amplitude was merely the observed change in flux.

In short, published phase amplitudes and phase offsets have self-consistent uncertainties, which suggests that including them improves the accuracy of our energy budgets.

\section{Apparent vs.\ Effective Temperatures}
\label{sec:appar_effec}
The view of a planet will affect the disk-integrated bolometric flux one measures, and hence the effective temperature one infers. Visibility is highest at the center of the planetary disk and lowest along the limb; the visibility is zero on the far side of the planet \citep{cowan2013light}. If a planet's hottest locations are directly facing the observer, then that hemisphere will likely appear hotter than its actual effective temperature, and vice versa.

Light curve inversion \citep{cowan2008inverting} provides a means to correct for longitudinal inhomogeneities in brightness and temperature, but this method has not been used to interpret most phase curves. Eclipse mapping can in principle constrain the meridional temperature gradients of the dayside, but so far it has only been applied to one planet in a single spectral band \citep{deWit_2012,Majeau_2012}.

The spatial inhomogeneity of short-period planets therefore presents a challenge to the analysis and interpretation of phase curves: strictly speaking, we need to know the hemispherical effective temperatures to constrain Bond albedo and day--night heat transport, but we can only measure apparent temperatures from a light curve. While apparent temperatures of exoplanets have been discussed before \citep[e.g.][]{Fortney_2006}, to our knowledge they have not been explicitly compared to effective temperatures.

For our analysis we use the semi-analytic energy balance model of \citet{cowan2011model}. In particular, we consider an idealized hot Jupiter that is on a circular orbit, has no internal heat, reflects no light, and has Eastward winds. 
We numerically solve for this planet's steady-state \emph{bolometric} flux on a grid in latitude and longitude. Next we assume an equatorial observer and integrate this flux two ways at each orbital phase. In one case, we include the observer's visibility of the planet and so calculate an apparent temperature for that hemisphere:
\begin{equation}
T_{\mathrm{app}} = \left(\frac{1}{\pi}\oint{V(\boldsymbol{\Omega
}) T^4(\boldsymbol{\Omega}) d\boldsymbol{\Omega}}\right)^\frac{1}{4},
\end{equation}
where $\boldsymbol{\Omega}$ denotes a location on the surface of the sphere.

In the other case, we use the full flux from every grid point that is visible at all, calculating the hemisphere's effective temperature: 
\begin{equation}
T_{\mathrm{eff}} = \left(\frac{1}{2\pi}\int{T^4(\boldsymbol{\Omega}) d\boldsymbol{\Omega}}\right)^\frac{1}{4},
\end{equation}
where the integral is only performed on the visible hemisphere of the planet.

\begin{figure*}
	\centering
	\includegraphics[width=0.625\linewidth]{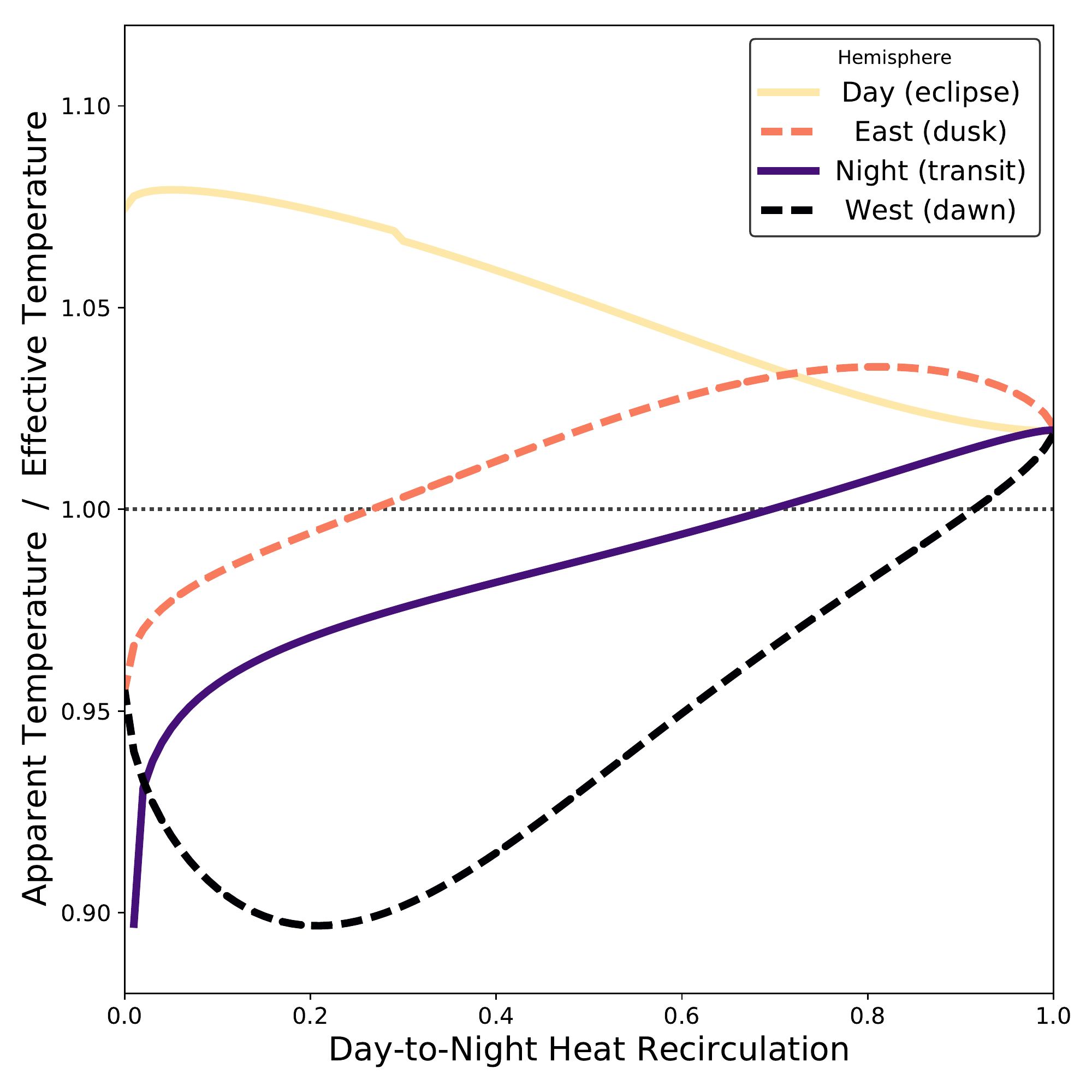}
	\caption{The ratio of apparent temperature (inferred from a \emph{bolometric} light curve) to effective temperature (based on total emitted flux) for four planetary hemispheres, as a function of recirculation. Here we use the energy balance model for planets on circular orbits in \citet{cowan2011model}. The dotted line shows where both temperatures are equal. By symmetry, the East and West hemispheres (dashed curves) have the same temperature ratio when $\varepsilon=0$. As expected, as $\varepsilon\rightarrow 1$ all curves approach an identical ratio: \mbox{$(32/3\pi^{2})^{1/4}\approx1.02$}. By estimating recirculation efficiency, one can empirically convert a planet's apparent temperatures into true effective temperatures.}
	\label{fig:apparenteffectivetratios}
\end{figure*}

Figure~\ref{fig:apparenteffectivetratios} shows the temperature ratio for four planetary hemispheres, or orbital phases, as a function of recirculation efficiency.\footnote{\citet{cowan2011model} defined recirculation efficiency, $\epsilon\in[0,\infty)$, as the product of a planet's radiative timescale and advective frequency, which differs from $\varepsilon\in[0,1]$ used by \citet{cowan2011statistics} and in this work. Testing our numerical flux grids, the best-fit function we find to convert between $\epsilon$ and $\varepsilon$ is:
	\begin{equation*}
	\label{eq:ep_varep}
	\varepsilon = \frac{\epsilon^{b}}{c + \epsilon^{b}},
	\end{equation*}
with $b=1.652$ and $c=1.828$. We use this equation in Figures~\ref{fig:JCS_OffsetvContrast_189_Tight} and \ref{fig:apparenteffectivetratios}.} Clearly, apparent and effective temperatures usually differ. At eclipse, a planet's hottest regions are almost always in view for an equatorial observer. We confirm that this leads one to overestimate the dayside effective temperature (yellow solid curve). Nightsides, on the other hand, appear overly cool for recirculation efficiencies up to $\approx0.7$ (dark solid curve). Together these results show that day and nightside effective temperatures are closer in value than their apparent temperatures. In fact, the real bolometric flux ratios of planets in Figure~\ref{fig:17Spring_Therm_MixedOff_Tight} would be $\sim20\%$ larger for HD~189733b to $\sim90\%$ for WASP-18b. Furthermore, with perfect recirculation the temperature ratio for all hemispheres converges to $\approx1.02$. This is inherent to our model because it does not include poleward heat transport.

If one combines the observed flux from two opposing hemispheres (i.e.\ values of the light curve at phases $180\degree$ apart), then one can try to estimate the effective temperature of the whole planet. In particular, we find the least-biased apparent temperatures are at phases between about $25\degree$ and $50\degree$ after transit and eclipse (not shown). 

We stress that these results are model-dependent. Nonetheless, Figure~\ref{fig:apparenteffectivetratios} suggests that na{\"i}vely combining apparent temperatures from two diametrically opposite hemispheres of a planet will generally yield the planet's global effective temperature to better than $10\%$. The scheme of \citet{cowan2011statistics}, essentially what we use in the current study, should perform considerably better: it is---by design---accurate in the limits of no heat transport and perfect heat transport.

\end{document}